\documentclass[12pt,epsfig,epsf]{article}
\usepackage{graphicx}
\usepackage{amssymb,amsbsy}
\usepackage{amsmath}

\def\be{\begin{equation}}
\def\ee{\end{equation}}
\def\bfi{\begin{figure}}
\def\efi{\end{figure}}
\def\bea{\begin{eqnarray}}
\def\eea{\end{eqnarray}}

\begin{document}

\begin{center}
{\Large \bf Crossover in Growth Law and Violation of Superuniversality in the Random Field Ising Model}
\vskip1cm
F. Corberi$^1$, E. Lippiello$^2$, A. Mukherjee$^3$, S. Puri$^3$ and M. Zannetti$^1$ \\
\vskip0.5cm
$^1$Dipartimento di Fisica {\it E.Caianiello} and CNISM, Unit\`a di Salerno,
Universit\`a di Salerno, via Ponte don Melillo, 84084 Fisciano (SA), Italy. \\
$^2$Dipartimento di Scienze Ambientali, Seconda Universit\`a di Napoli, \\
Via Vivaldi, Caserta, Italy. \\
$^3$School of Physical Sciences, Jawaharlal Nehru University, \\
New Delhi--110067, India. \\
\end{center}

\begin{abstract}
We study the nonconserved phase ordering dynamics of the $d=2,3$ random field Ising model, quenched 
to below the critical temperature. 
Motivated by the puzzling results of previous work in two and three dimensions, reporting a crossover
from power-law to logarithmic growth, {\it together} with superuniversal behavior of the correlation function, 
we have undertaken a careful investigation of both the domain growth law and the autocorrelation function. 
Our main results are as follows: We confirm the crossover to asymptotic logarithmic behavior in the growth law,
but, at variance with previous findings, the exponent in the preasymptotic power law
is {\it disorder-dependent}, rather than being the one of the pure system. Furthermore, we find that
the autocorrelation function {\it does not} display superuniversal behavior. 
This restores consistency with previous results for the $d=1$ system, and fits nicely into the unifying scaling 
scheme we have recently proposed in the study of the random bond Ising model.

\end{abstract}

\newpage

\section{Introduction}
\label{s1}

Much recent interest in statistical physics has focused on understanding
out-of-equilibrium phenomena. In this context, of paramount importance are slow relaxation
phenomena, which primarily occur in glassy systems. An important hallmark of slow relaxation is
the lack of time-translation-invariance, manifested through aging behavior. A similar
phenomenology is also observed in systems without disorder, e.g., 
ferromagnets quenched below the critical point. The behavior of these systems is well understood
in terms of the domain growth mechanism of slow relaxation \cite{Bray,Puri}. 

The key feature of domain growth, or coarsening, is the unbounded growth of the domain size,
which entails scaling due to the existence of a dominant length scale, and aging as a
manifestation of scaling in multiple-time observables.
The simplicity of this structure is very attractive and is expected to be valid beyond the 
realm of disorder-free phase-separating systems, 
establishing domain growth as a paradigm of slow relaxation. 
However, as is well known, the applicability of domain growth concepts to {\it hard}
problems (such as spin glasses or structural glasses) still remains a debated issue \cite{reviews}. 
Therefore, it is of
considerable interest to study the role of disorder in systems where its presence 
does not compete with phase ordering~\cite{sp04}. 

A class of systems of this type are disordered ferromagnets, where disorder
coexists with the low-temperature ferromagnetic order. There are different ways to introduce
disorder in a ferromagnet without inducing frustration. This can be achieved through bond or site
dilution, by randomizing the exchange interaction strength while keeping it ferromagnetic, or
by introducing a random external field. These disordered systems have been an active area of research
for quite some time now. The unifying theme of investigation has been disorder-induced changes in
the properties of the underlying pure systems, with primary 
interest in the {\it growth law}, in the {\it equal-time correlation function} and, more recently, 
in the {\it two-time autocorrelation function} and the related {\it response function} \cite{HP1,HP2,EPL,Park,CLMPZ}.
However, despite the many experimental and theoretical studies \cite{sp04}, 
a number of issues are still open. Among these, of primary importance are (i) the nature of the
asymptotic growth law ({\it power-law} vs. {\it logarithmic}) and (ii) the existence of
{\it superuniversal} behavior of the correlation and response functions. This is the idea
that scaling functions are robust with respect to disorder, which is expected not to change the low temperature 
properties of the system \cite{Cuglia}.  
The lack of a general framework to understand this complex phenomenology has proven a major obstacle
to development. Recently, in the context of the random bond Ising model (RBIM)~\cite{CLMPZ}, 
we have shown that the renormalization group (RG) picture of crossover phenomena may well serve the
purpose.

In this paper, we extend the RG conceptual framework to
the ordering dynamics of the random field Ising model (RFIM) \cite{nattermann}. 
In this system, the deep asymptotic regime turns out to be numerically accessible,
allowing us to make precise statements regarding the growth law and the
superuniversality (SU) issue (see subsection~\ref{SU}). Our principal findings are (a) the existence of a
crossover from power law domain growth (with a {\it disorder-dependent} exponent)
to logarithmic growth, and (b) the absence of SU.
Both results fit nicely into an RG picture where disorder acts as a relevant perturbation with respect to 
the pure fixed point. This confirms the robustness and the general applicability
of the approach proposed in Ref.~\cite{CLMPZ}.

This paper is organized as follows. In Sec.~\ref{s2}, we provide an overview of domain growth laws 
and of our scaling framework for 
phase ordering dynamics in disordered systems. In Sec.~\ref{s4}, we present 
detailed numerical results for ordering in the $d=2$ RFIM\footnote{As it is explained in subsection~\ref{SD},
the spin updating rule we use is equivalent to a quench to $T=0$. Hence, the $d=2$ RFIM
phase orders even if $d=2$ is the lower critical dimensionality.}. These results are 
interpreted using the scaling framework of Sec.~\ref{s2}. Sec.~\ref{s5} is devoted
to the presentation of numerical results in the $d=3$ case. Finally, in Sec.~\ref{s6}, 
we conclude this paper with a summary and discussion.

\section{Domain Growth Laws in Disordered Systems}
\label{s2}

Dynamical scaling is the most important characteristic of phase ordering systems~\cite{Bray,Puri}.
Let us summarise the concept in the simplest case of pure systems. 
As time increases, the typical domain size $L(t)$ grows and becomes
the dominant length in the problem. Then, all other lengths can be rescaled with respect to $L(t)$.
For instance, the two-time order-parameter correlation function $C(r,t,t_w)$, with $t_w \leq t$, 
scales as~\cite{furukawa,chpt}
\be
C(r,t,t_w) = G(r/L,L/L_w)
\label{SCL.1}
\ee
where $L$ and $L_w$ stand for $L(t)$ and $L(t_w)$, respectively.
This contains, as a special case, the usual scaling of the equal-time ($t=t_w$) correlation 
function $C(r,t)=G_1(r/L)$. Further, for $r=0$, we have the aging form of the autocorrelation 
function $C(t,t_w) = G_2(L/L_w)$.
The validity of scaling is, by now, a well-established fact. A complete picture of an ordering problem 
requires the understanding of the growth law (i.e. how $L(t)$ depends on $t$) and of the scaling 
function $G(x,y)$.

A systematic study of the growth law has been undertaken by Lai et al. (LMV)~\cite{Lai,Tafa}, who identified
four different universality classes of growth kinetics. LMV considered the role of several factors 
in the ordering
dynamics, e.g., temperature, conservation laws, dimensionality, order parameter symmetry, 
lattice structure and disorder. An important distinction is made between systems that do not freeze 
(i.e., without free-energy barriers) and those that do freeze (i.e., with barriers) when the quench 
is made to $T=0$. To the first category belong pure systems with non-conserved dynamics, whose growth 
follows the power law $L(t) = Dt^{1/z}$, where $z=2$. LMV designate these
as {\it Class 1} systems. To the second category belong systems whose growth requires thermal activation. 
This includes pure systems with conserved order parameter and systems (both conserved and non-conserved) 
with quenched disorder. This category is further subdivided into three classes. In {\it Class 2} systems the 
freezing involves only local defects, with activation energy $E_B$ independent of the domain size. In this
case, growth is still power law: $L(t)=Dt^{1/z}$ with $z=2$ and 3 for the non-conserved and conserved cases,
respectively. Furthermore, the prefactor $D$ has a strong temperature dependence, $D \sim e^{-E_B/(zT)}$. 
Finally, in {\it Class 3} and {\it Class 4} systems, the freezing involves a collective behavior 
which depends on the domain size $L$. If the corresponding activation energy scales with
$L$ like $E_B(L)\sim \epsilon L^{\varphi}$, 
where $\epsilon$ measures the disorder strength, the asymptotic growth law is logarithmic
\be
L(t) \sim (T/\epsilon)^{1/\varphi} \left[ \ln (t/\tau)\right]^{1/\varphi}
\label{class}
\ee
with $\tau \sim T/(\varphi \epsilon)$. 
For {\it Class 3} systems, we have $\varphi =1$, and for {\it Class 4} systems, we have $\varphi \neq 1$.

Ferromagnets (with or without disorder) offer examples of the classes listed above. For simplicity, 
let us consider systems with non-conserved order parameter. The pure ferromagnetic Ising 
model with Glauber kinetics is a well-known Class 1 system~\cite{Bray}. The $d=1$ ferromagnetic RBIM~\cite{EPL} 
is an example of a Class 2 system. The $d=1$ RFIM~\cite{decandia} belongs 
to Class 4, with $\varphi = 1/2$. The RFIM in higher dimensions, $d=2$~\cite{pp93,eo90} and 
$d=3$~\cite{rao,aron}, shows logarithmic growth, although it is not easy to unambiguously establish the 
value of $\varphi$. Recently, we have presented evidence~\cite{CLMPZ} for logarithmic growth in
the $d=2$ RBIM, but have not established whether it is a Class 3 or Class 4 system. This is a 
particularly interesting system, because its growth law was previously~\cite{ppr04} believed to be power law with 
a disorder-dependent exponent. If so, this would have shown the existence of a new universality class, say {\it Class 5}, 
in addition to the 
four listed by LMV. We should stress that a huge numerical effort is involved in accessing the logarithmic
growth regime of the $d=2$ RBIM, and our understanding of this system remains incomplete.  

In Ref.~\cite{CLMPZ}, we have proposed to unify this wide variety of behaviors for disordered domain growth 
into a scaling framework for the growth law itself. In all the cases we consider in this paper, disorder ($h_0$) 
and temperature ($T$) enter through their ratio $h_0/T$ (see Sec. \ref{SD} below). 
This will be denoted by $\epsilon$ and, for 
short, will be termed as disorder. Let us begin with the straightforward crossover set-up,
where the growth law is assumed to scale as
\be
L(t,\epsilon) = t^{1/z} F(\epsilon/t^{\phi}),
\label{SH.1}
\ee
$z=2$ is the growth exponent for non-conserved dynamics in a pure ferromagnet and $\phi$ is the
crossover exponent. With the additional assumption that the scaling function behaves as
\be
F(x)  \sim \left \{ \begin{array}{ll}
        $const.$ ,\;\; $for$ \;\; x \rightarrow 0,\\
        x^{1/(\phi z)} \ell ( x^{-1/\phi})  ,\;\; $for$ \;\; x \rightarrow \infty,
        \end{array}
        \right .
        \label{SH.1bis}
        \ee
where $x=\epsilon/t^{\phi}$, Eq.~(\ref{SH.1}) describes the crossover from 
the power law $L(t) \sim
t^{1/z}$ to the asymptotic form $L(t) \sim \ell \left( t/\epsilon^{1/\phi} \right)$,
if $\phi < 0$, and vice-versa if $\phi > 0$. Alternatively, disorder is asymptotically relevant when $\phi < 0$, 
and irrelevant when $\phi > 0$. 
The key quantity in the analysis of crossover is the effective growth exponent 
\be
{1 \over z_{\rm eff}(t,\epsilon)} = {\partial \ln L(t,\epsilon) \over \partial \ln t}
= {1 \over z} - \phi {\partial \ln F(x) \over \partial \ln x},
\label{ND.1}
\ee
which depends on $t$ and $\epsilon$ through $x$.

In the following discussion, it will be useful to use the above relations in the inverted form:
\be
t= L^z g(L/\lambda),
\label{SH.2}
\ee
where
\be
\lambda = \epsilon^{1/(\phi z)}
\label{SH.3}
\ee
is a length scale associated with disorder. The scaling functions appearing in Eqs.~(\ref{SH.1}) 
and~(\ref{SH.2}) are related by
\be
g(y) = F^{-z}(x)
\label{SH.4}
\ee
and $y=L/\lambda$ is related to $x$ by
\be
y=x^{-1/(\phi z)} F(x).
\label{SH.5}
\ee
Then, from Eq.~(\ref{SH.1bis}) and $\phi < 0$, it follows that
\be
g(y)  \sim \left \{ \begin{array}{ll}
        $const.$  ,\;\; $for$ \;\; y \ll 1,\\
        y^{-z}\ell^{-1}(y) ,\;\; $for$ \;\; y \gg 1,
        \end{array}
        \right .
        \label{SH.7}
        \ee
where $\ell^{-1}$ stands for the inverse function of $\ell$. The opposite behavior holds for $\phi > 0$
\be
g(y)  \sim \left \{ \begin{array}{ll}
        y^{-z}\ell^{-1}(y) ,\;\; $for$ \;\; y \ll 1,\\
        $const.$ ,\;\; $for$ \;\; y \gg 1.
        \end{array}
        \right .
        \label{SH.6}
        \ee
Finally, the effective exponent as a function of $y$ is obtained from Eq.~(\ref{SH.2})
\be
z_{\rm eff}(y) = z + {\partial \ln g(y) \over \partial \ln y}.
\label{SH.8}
\ee
Therefore, for $\phi < 0$ (disorder relevant), Eq.~(\ref{SH.7}) yields
\be
z_{\rm eff}(y)  = \left \{ \begin{array}{ll}
         z ,\;\; $for$ \;\; y \ll 1,\\
        \partial \ln \ell^{-1}(y)/\partial \ln y ,\;\; $for$ \;\; y \gg 1,
        \end{array}
        \right .
        \label{SH.10}
        \ee
and for $\phi > 0$ (disorder irrelevant), we obtain from Eq.~(\ref{SH.6}) 
\be
z_{\rm eff}(y)  = \left \{ \begin{array}{ll}
        \partial \ln \ell^{-1}(y)/\partial \ln y ,\;\; $for$ \;\; y \ll 1,\\
        z ,\;\; $for$ \;\; y \gg 1.
        \end{array}
        \right .
        \label{SH.9}
        \ee

\subsection{Superuniversality}
\label{SU}

One would expect that the above crossover scenario, which is well established for the growth law,
would extend also to the other observables. However, this expectation is in conflict with
the SU statement that all disorder dependence in
observables other than the growth law can be
eliminated by reparametrization of time through $L(t,\epsilon)$ \cite{Cuglia}. Thus, according to SU,
for the autocorrelation function one should have
\be
C(t,t_w,\epsilon) = G_2(L(t_w,\epsilon)/L(t_w,\epsilon))
\label{sup.1}
\ee
where $G_2$ is the scaling function of the pure case. The validity of SU is controversial, since
the $d=1$ results~\cite{EPL,decandia} clearly demonstrate the absence of SU, while
from the study of the correlation function for $d \geq 2$, there is evidence both
in favour~\cite{rao,aron,ppr04,pcp91} and against \cite{CLMPZ} SU validity.
Recently, the validity of SU has been extended to the geometrical properties of domain structures \cite{Sicilia}.
 
In the next sections we present comprehensive numerical results from large scale simulations of 
ordering dynamics in the RFIM in $d=2,3$. We will analyze numerical results within
the above scaling framework, producing evidence against SU validity.

\section{Numerical Results for $d=2$}
\label{s4}

\subsection{Simulation Details}
\label{SD}

We consider an RFIM on a two-dimensional square lattice, with the Hamiltonian 
\be
H= -J\sum_{\langle ij \rangle}\sigma_i\sigma_j - \sum_{i=1}^{N} h_i\sigma_i,~~~~~~~\sigma=\pm1,
\label{SCL.9}
\ee
where $\langle ij \rangle$ denotes a nearest-neighbour pair, and $J > 0$ is the ferromagnetic 
exchange coupling.
The random field $h_i= \pm h_0$ is an uncorrelated quenched variable with a bimodal distribution
\be
P(h_i)=\frac{1}{2} \left[\delta(h_i-h_0)+\delta(h_i+h_0) \right] .
\ee

The system evolves according to the Glauber  kinetics, which models
nonconserved dynamics \cite{Puri}, with spin flip transition rates given by
\be
w(\sigma_i \rightarrow -\sigma_i)=\frac{1}{2}\left (1-\sigma_i \tanh 
\left ( (H_i^W+h_i)/T\right) \right)
\ee
where $H_i^W$ is the local Weiss field.
All results in this paper correspond to the limit  $T \rightarrow 0$
($J/T \rightarrow \infty$), while keeping
the ratio $\epsilon = h_0/T$ finite. In this limit the system undergoes phase ordering 
in any dimension, down to $d=1$ \cite{decandia}.
The transition rates take the form
\be
w(\sigma_i \rightarrow -\sigma_i) = 
\left \{ \begin{array}{ll}
        1\;\; $for$ \;\; H_i^W\sigma_i <0 ,\\
        0\;\; $for$ \;\; H_i^W\sigma_i>0 ,\\
         \frac{1}{2}\left (1-sign(\sigma_i h_i) \tanh (\epsilon) \right) 
        \;\; $for$ \;\; H_i^W=0
        \end{array}
        \right .
        \label{wT0}
        \ee
which shows that disorder affects the evolution 
through the ratio $\epsilon=h_0/T$, as anticipated in Sec.~\ref{s2}.
Moreover, Eq.~(\ref{wT0}) allows for an accelerated updating rule, with a
considerable increase in the speed of computation \cite{nobulk},
by restricting updates to the sites with $H_i^W \sigma_i\le 0$, whose number decreases in time as $1/L(t)$. 
The gain in the speed of computation becomes more important the longer the
simulation. 

All statistical quantities presented here have been obtained as an average 
over $N_{\rm run}=10$ independent runs. For each run, the system has  
different initial condition and random field configuration. 
We have considered the values of disorder amplitude $\epsilon=0,0.25,
0.5,1,1.5,2,2.5$ and we have
carefully checked that no finite size effects are present 
up to the final simulation time when $N=8000^2$ spins. In the pure case,  
since coarsening is more rapid, we have taken $N=12000^2$.

Numerical results for the growth law and the autocorrelation function are presented in 
the following subsections.

\subsection{Growth Law}

\begin{figure}
   \centering
  \rotatebox{-90}{\resizebox{.85\textwidth}{!}{\includegraphics{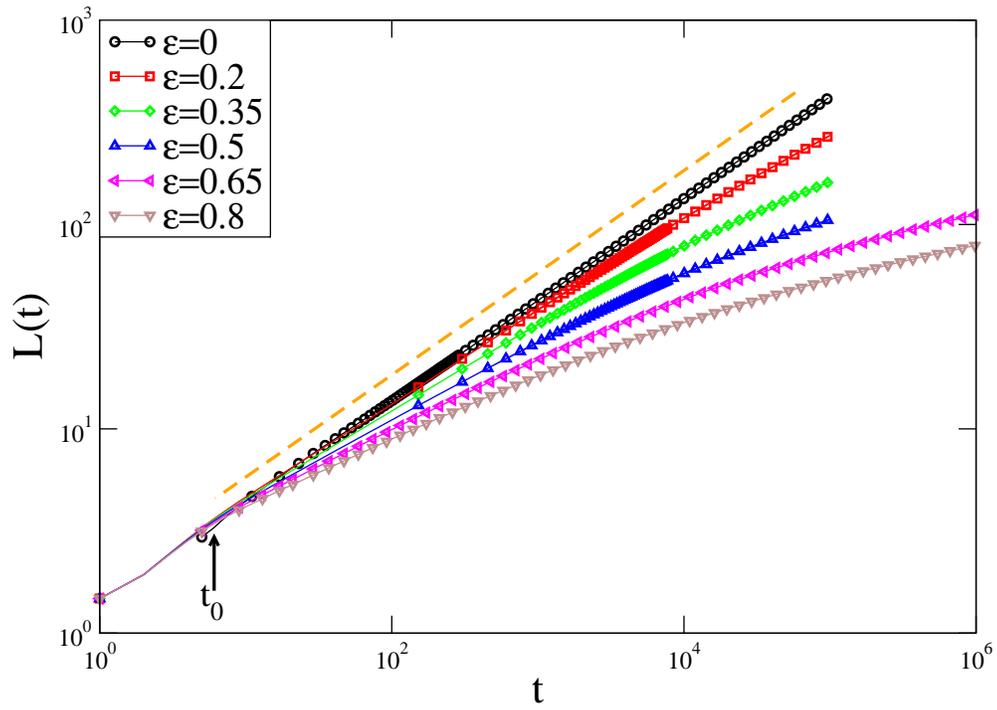}}}
  \vspace{0.5cm}
   \caption{(Color online) Growth law in $d=2$. The dashed line is the $t^{1/2}$ growth law.}
\vspace{1cm}
\label{fig1}
\end{figure}

We have obtained the characteristic $L(t)$ from the inverse density of defects. This is measured by 
dividing the number of sites with at least one oppositely-aligned neighbor by the total number of 
sites\footnote{We have checked that the same results are obtained measuring $L(t)$ from the
equal time corrlation function.}.
The plot of $L(t,\epsilon)$ vs. $t$, in Fig.~\ref{fig1}, shows the existence 
of at least two time-regimes, separated by a microscopic time $t_0$ of order 1. In the early-time regime 
(for $t < t_0$), there is no dependence on disorder and growth is fast. This is the regime where the 
defects seeded by the random initial condition execute rapid 
motion toward the nearby local minima. For $\epsilon > 0$ and $t > t_0$, there is a strong
dependence on disorder producing slower growth and deviation from the power law behavior of 
the pure case (top cicles line in Fig.~\ref{fig1}).

\begin{figure}
\centering
\includegraphics[width=0.8\textwidth]{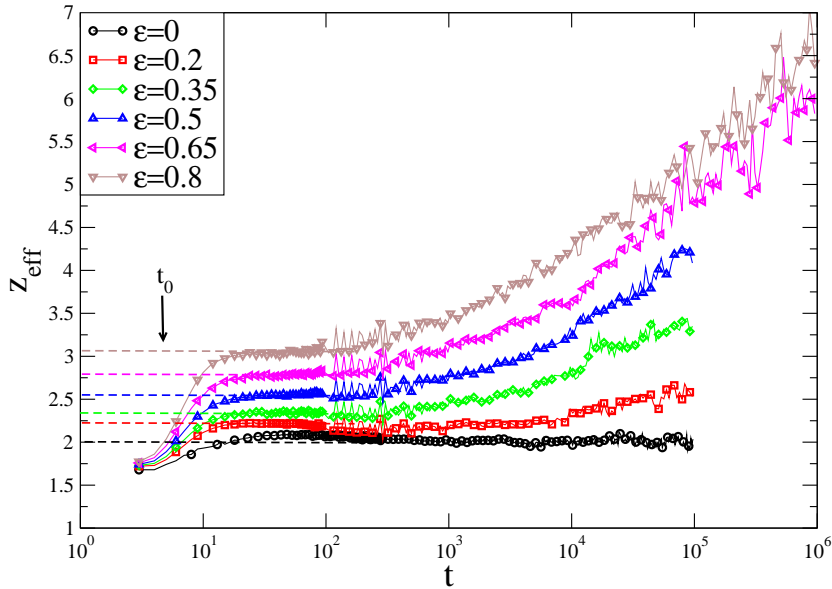}
\caption{(Color online) Effective exponent $z_{\rm eff}$ vs. $t$ in $d=2$. 
The horizontal dashed lines indicate $\overline{z}$, 
the plateau values of $z_{\rm eff}$.}
\vspace{1cm}
\label{fig2}
\end{figure}

\begin{figure}
\centering
\includegraphics[width=0.6\textwidth]{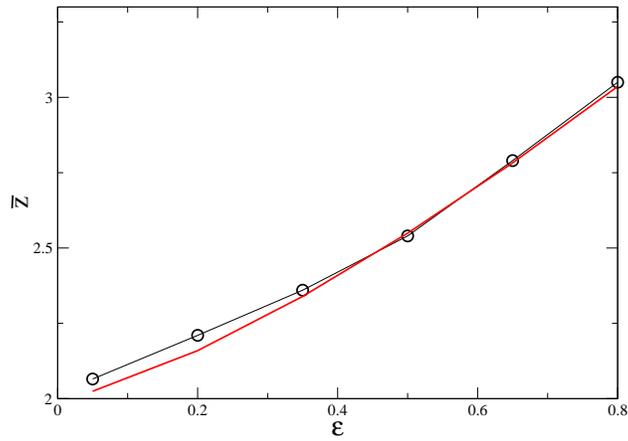}
\caption{(Color online) $\overline{z}$ (taken from Fig.~\ref{fig2}) vs. $\epsilon$. 
The red line is the best fit $\overline{z} = 2.0+1.4 \epsilon^{1.35}$.}
\vspace{1cm}
\label{fig3}
\end{figure}

In Fig.~\ref{fig2}, we show the time-dependence of the effective exponent $z_{\rm eff}(t,\epsilon)$, 
defined by Eq.~(\ref{ND.1}). For $t > t_0$, this plot shows the existence of an intermediate power-law regime,
characterized by a plateau where $z_{\rm eff}$ is approximately constant. 
This is followed by the late regime where $z_{\rm eff}$ is clearly time-dependent. 
The disorder dependent values of $z_{\rm eff}$ on 
the plateaus, denoted by $\overline{z}$, are listed in Table~\ref{expon} and plotted in Fig.~\ref{fig3}.
We encountered a similar crossover in our study of the $d=2$ RBIM~\cite{CLMPZ}, i.e., 
a preasymptotic power law regime with
a disorder dependent exponent, followed by an asymptotic regime where the growth law deviates from a power law.
\begin{table}
\begin{center}
\begin{tabular}{|c|c|}   \hline
$\epsilon$  & $\overline{z}$ \\  \hline
$0$ & $2$ \\         \hline
$0.2$ & $2.20$ \\    \hline
$0.35$ & $2.31$ \\    \hline
$0.5$ & $2.52$ \\    \hline
$0.65$ & $2.77$  \\    \hline
$0.8$ & $3.05$  \\    \hline
\end{tabular}
\end{center}
\caption{Plateau exponent $\overline{z}$ for various disorder strengths.}
\vspace{1cm}
\label{expon}
\end{table}

\begin{figure}
\centering
\rotatebox{0}{\resizebox{.85\textwidth}{!}{\includegraphics{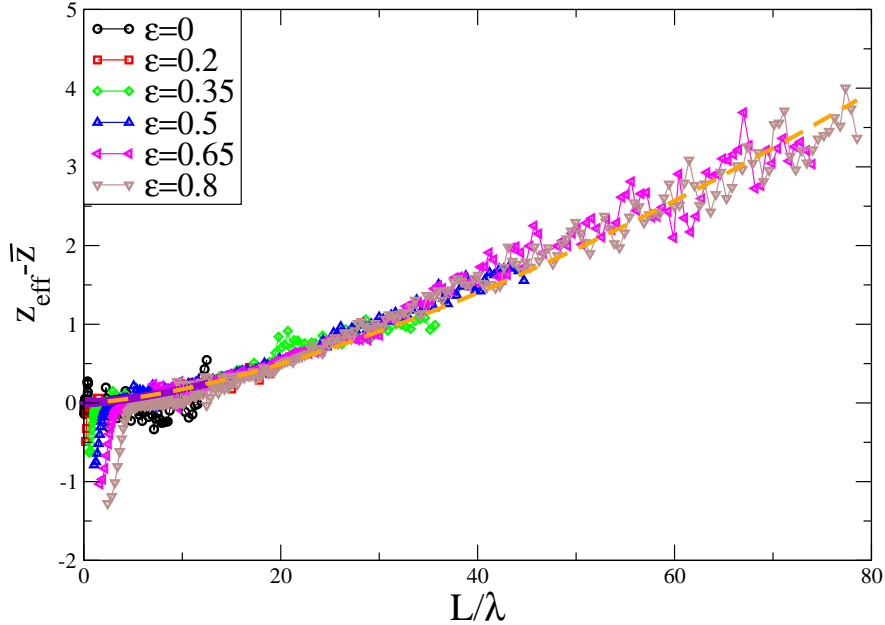}}}
\vspace{0.5cm}
\caption{(Color online) Subtracted effective exponent $(z_{\rm eff}-\overline{z})$ vs. $L/\lambda$.
The dashed line is the best fit $z_{\rm eff}-\overline{z}= 0.0055(L/\lambda)^{1.5}$.}
 \vspace{1cm}
\label{fig4}
\end{figure}

\begin{figure}
\centering
\includegraphics[width=0.7\textwidth]{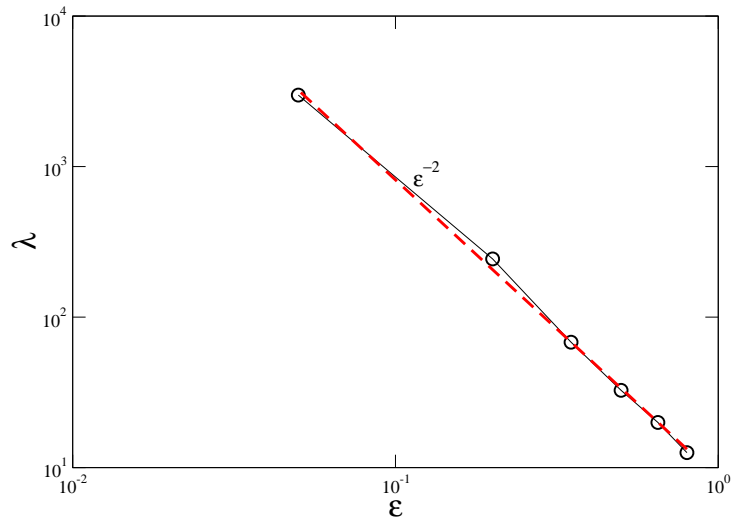}
\caption{(Color online) Plot of $\lambda$ vs. $\epsilon$.}
\vspace{1cm}
\label{fig5}
\end{figure}

The appearance of a disorder dependent exponent $\overline{z}$ in the intermediate regime
suggests to upgrade the crossover picture, presented in Sec.~\ref{s2}, by
replacing the pure growth exponent $z$ by $\overline{z}$ in all the scaling formulae.
Then, from Eq.~(\ref{SH.8}) it follows that $z_{\rm eff}-\overline{z}$ 
ought to depend only on $y = L/\lambda$. Indeed, as Fig.~\ref{fig4} shows, it is possible
to determine numerically the quantity $\lambda$ such that 
the plots of $(z_{\rm eff}-\overline{z})$ vs. $L/\lambda$, for different disorder values, collapse 
on a single master curve.  The $\epsilon$-dependence of $\lambda$  
is displayed in Fig.~\ref{fig5} and is well fitted by
\be
\lambda \sim \epsilon^{-2}.
\label{ND.1bis}
\ee
Comparing this with Eq.~(\ref{SH.3}), the negative exponent implies $\phi < 0$ and, therefore,
that disorder acts like a relevant scaling field. This is also confirmed by the behavior 
of $z_{\rm eff}(y)$ in Fig.~\ref{fig4}, which is consistent with Eq.~(\ref{SH.10}) 
but not with Eq.~(\ref{SH.9}).

\begin{figure}
\centering
\rotatebox{0}{\resizebox{.85\textwidth}{!}{\includegraphics{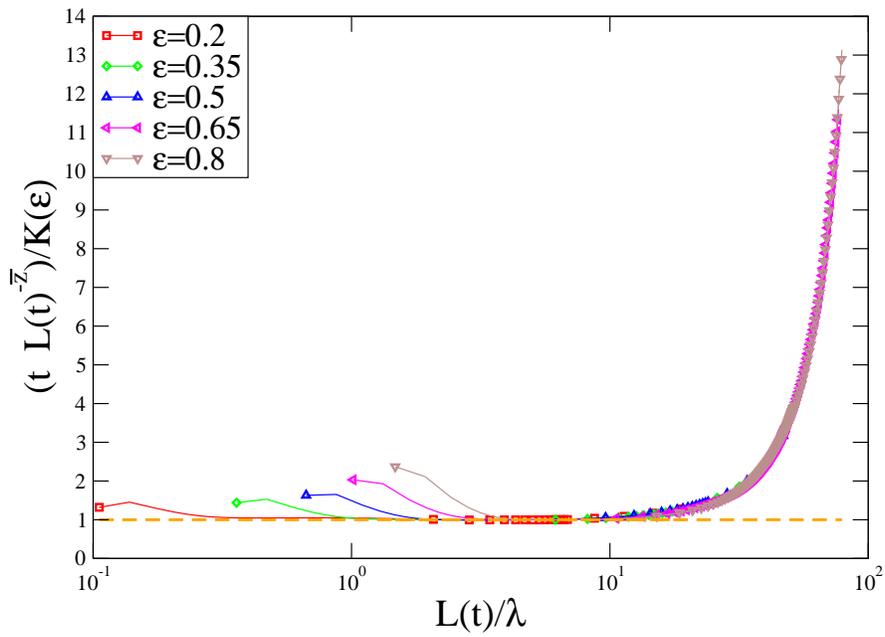}}}
\caption{(Color online) Plot of $t L^{-\overline{z}}/K(\epsilon)$ vs. $L/\lambda$ with various disorder values.
The master curve obeys the exponential form of Eq.~(\ref{num.6}).}
\vspace{1cm}
\label{fig6}
\end{figure}

Fitting the data of Fig.~\ref{fig4} to the power law $z_{\rm eff} - \overline{z} = by^{\varphi}$, we
find $b \simeq 0.0055$ and $\varphi \simeq 1.5$. Hence, from the definition of
$z_{\rm eff}$ in Eq.~(\ref{ND.1}) follows
\be
{\partial \ln t \over \partial L} = \overline{z} + by^{\varphi}
\label{xx.1}
\ee
which, after integrating with respect to $L$, yields
\be
t = K(\epsilon)L^{\overline{z}} g(L/\lambda)
\label{num.4}
\ee
where $K(\epsilon)$ is an $\epsilon$-dependent prefactor. Indeed, replotting in Fig.~\ref{fig6}
the data of Fig.~\ref{fig1}, as $t L^{-\overline{z}} /K(\epsilon)$ vs. $y$, an excellent data
collapse on the master curve 
\be
g(y) \sim \exp\left( \frac{b}{\varphi}y^{\varphi} \right)
\label{num.6}
\ee
is obtained, with the values of $K(\epsilon)$ listed in Table~\ref{prefactor}.

\begin{table}
\begin{center}
\begin{tabular}{|c|c|}   \hline
$\epsilon$  & $K$ \\  \hline
$0.5$ & $0.39$ \\         \hline
$1.0$ & $0.13$ \\    \hline
$1.5$ & $0.039$ \\    \hline
$2.0$ & $0.015$ \\    \hline
\end{tabular}
\end{center}
\caption{Prefactor $K(\epsilon)$ for various disorder strengths.}
\label{prefactor}
\end{table}
The plot of the scaling function $g(y)$
illustrates quite effectively (i) the existence of the crossover, and (ii) 
that our numerical data reach deep into the asymptotic regime. The flat part of the curve, 
where $g(y)$ lies on the horizontal dashed line at $g(y)=1$, corresponds to the preasymptotic 
power-law regime [cf. Eq.~(\ref{SH.7})]. The sharp and fast increase of $g(y)$, for large $y$, 
corresponds to the crossover to the asymptotic growth law
\be
{L \over \lambda} \simeq \left [ {\varphi \over b} \ln \left( t/\lambda^{\overline{z}} \right) \right ]^{1/\varphi}
\label{num.6bis}
\ee
which corresponds to the {\it Class 4} form of Eq.~(\ref{class}).

Summarising, our main findings for the growth law, in the $d=2$, case are as follows:

\begin{enumerate}

\item Disorder is a relevant perturbation with respect to pure-like behavior.

\item The corresponding growth law shows a clear crossover from power-law to logarithmic behavior:
\be
L(t,\epsilon)   \sim \left \{ \begin{array}{ll}
        t^{1/\overline{z}} ,\;\; $if$ \;\; L \ll L_{\rm cr},\\
        (\ln t)^{1/\varphi} ,\;\; $if$ \;\; L \gg L_{\rm cr}.
        \end{array}
        \right .
        \label{num.7}
        \ee
This differs from previously found results, since the preasymptotic power law is not pure-like, 
due to the $\epsilon$-dependence of the exponent $\overline{z}$. This feature, also observed in 
the $d=2$ RBIM~\cite{CLMPZ}, means that disorder although globally relevant acts like a marginal 
operator in the neighborhood of the pure fixed point \cite{wegner}.

\end{enumerate}

Finally, we remark on the considerable numerical advanyage in using the effective exponent as a probe for the crossover. 
In fact, while the switch from preasymptotic to asymptotic behaviors in $z_{\rm eff}$ takes 
place at about $L_{\rm cr} \simeq \lambda$, from Eqs.~(\ref{num.4}) and~(\ref{num.6}), 
it follows that the condition $b y^{\varphi}/ \varphi  = 1$ puts the crossover, when looking at the 
domain size, at 
the much greater value $L_{\rm cr} \simeq 50 \lambda$, as it is evident from Fig.~\ref{fig6}.

\subsection{Autocorrelation Function and SU Violation}

The results presented above show that disorder affects the growth law as an 
asymptotically relevant parameter. 
Therefore, one would expect this to apply also to other observables. However,
as explained in Sec.~\ref{SU}, such an expectation would be in conflict with 
claims of SU validity.

\begin{figure}
\centering
\rotatebox{0}{\resizebox{.85\textwidth}{!}{\includegraphics{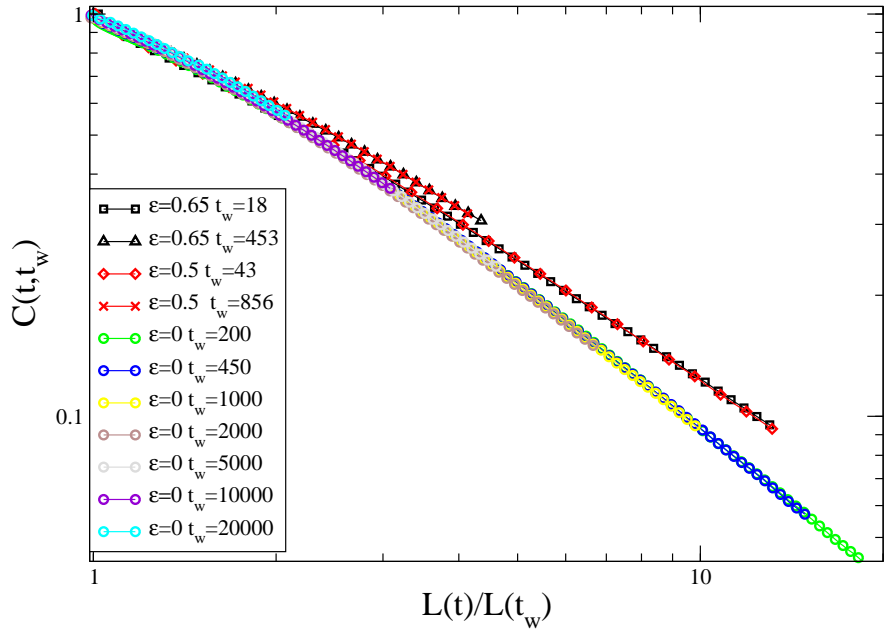}}}
\caption{(Color online) Autocorrelation function in $d=2$,
for disorder values and waiting times $t_w$
chosen so that $v=L_w/\lambda$ takes the three values $v =0,0.25,0.85$ corresponding, from bottom to top,
to the three different master curves.}
\vspace{1cm}
\label{fig7}
\end{figure}

\begin{figure}
\centering
\rotatebox{0}{\resizebox{.85\textwidth}{!}{\includegraphics{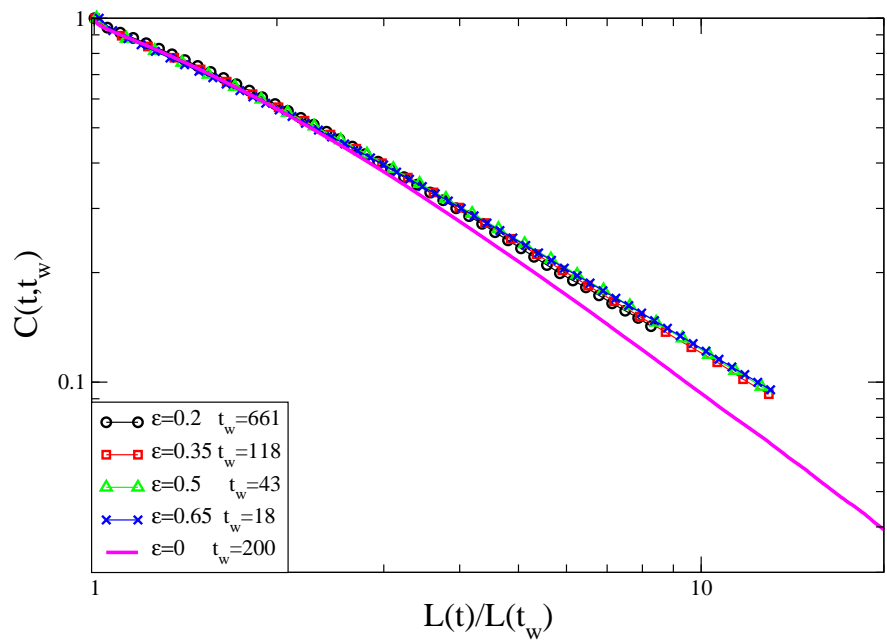}}}
\caption{(Color online) Analogous to Fig.~\ref{fig7}, with $v = 0$ (lower master curve) and $v=0.25$ 
(upper master curve).}
\vspace{1cm}
\label{fig8}
\end{figure}

In this section, 
we study the autocorrelation function, defined by
\be
C(t,t_w,\epsilon) = \langle \sigma_i(t) \sigma_i(t_w) \rangle
\label{SU.1}
\ee
which is independent of $i$, due to space-translation invariance.   
In Fig.~\ref{fig7} $C(t,t_w,\epsilon)$ has been plotted against $L/L_w$, for $\epsilon=0,0.5,0.65$
and with different values of $t_w$, chosen in such a way that the ratio $v=L_w/\lambda$
takes the three different values $v=0,0.25,0.85$.
If SU were valid all the curves, irrespective of the value of $\epsilon$, should collapse
on the $\epsilon=0$, or $v=0$, master curve. Instead, there is an evident $\epsilon$-dependence which
excludes SU validity. In addition, curves with the same value of $v$ do collapse, showing that
the autocorrelation function obeys the extended aging form
\be
C(t,t_w,\epsilon) = h \left ({L  \over L_w}, {\lambda \over L_w} \right ). 
\label{SU.2}
\ee
The pure master curve, with $v=0$, lies below the two corresponding to 
$v=0.25$ and $v=0.85$. Figure~\ref{fig8} displays a similar plot, with $v=0$ and $v=0.25$. The latter value is
obtained by combining the four different values of disorder $\epsilon=0.2,0.35,0.5,0.65$
with appropriately chosen $t_w$-values. Again, there are two distinct master curves for different $v$-values.

The reported SU violation is in agreement with the behavior of the autocorrelation function 
in the $d=1$ RFIM~\cite{decandia}.

\section{Numerical Results for $d=3$}
\label{s5}

As mentioned above, in previous studies of the $d=3$ RFIM \cite{rao,aron} SU has been found to hold. 
Here, instead, we present new results for this system, which produce evidence for
the same pattern of SU violation observed in the $d=2$ case.

Simulations were made on a system with $N=300^3$ spins on a cubic lattice, evolving with the transition
rates~(\ref{wT0}) and averaging over $N_{\rm run}=20$ runs, each 
with different initial condition and random field configuration.
We have considered the three disorder values $\epsilon=0.5,1,2$.
Although the quality of the data does not allow for an analysis of high precision
as in the $d=2$ case discussed above, nonetheless the main features, including the lack of SU validity,
do emerge quite clearly.

\begin{figure}
\centering
\rotatebox{0}{\resizebox{.85\textwidth}{!}{\includegraphics{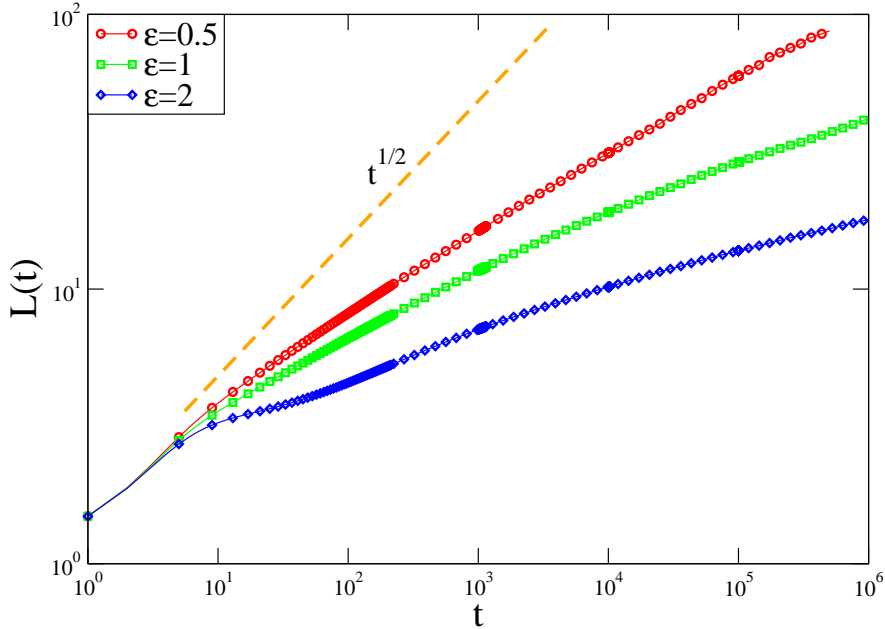}}}
\caption{(Color online) Growth law in $d=3$ for different disorder values.}
\vspace{1cm}
\label{fig9}
\end{figure}

\begin{figure}
\centering
\rotatebox{0}{\resizebox{.85\textwidth}{!}{\includegraphics{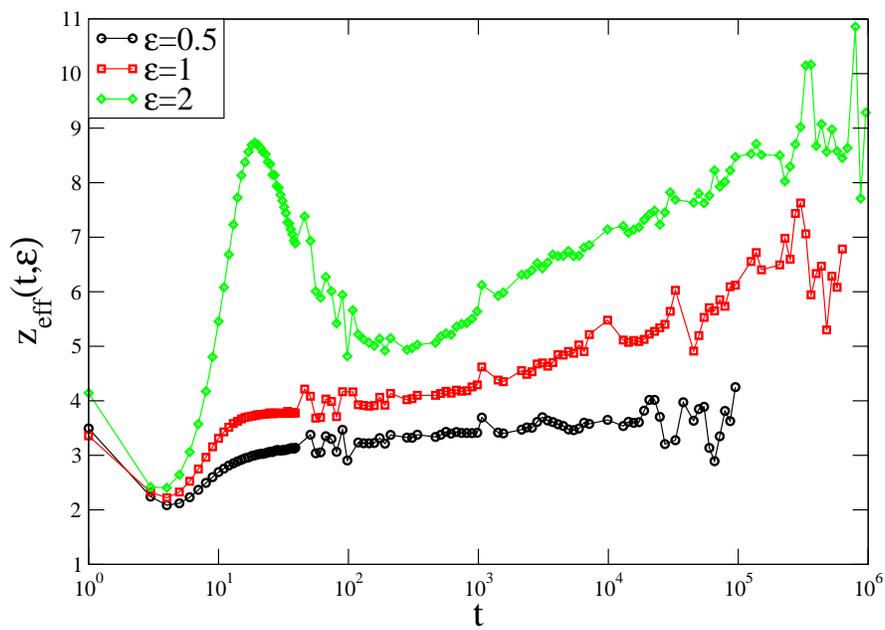}}}
\caption{(Color online) Effective exponent $z_{\rm eff}(t,\epsilon)$ in $d=3$.}
\vspace{1cm}
\label{fig10}
\end{figure}

Let us begin with the growth law. The $L(t)$ data have been plotted in Fig. \ref{fig9}.
The qualitative behaviour is  the same as
in Fig. \ref{fig1}, namely, as disorder increases, growth slows down. 
The corresponding effective exponent $z_{\rm eff}$ is displayed in Fig. \ref{fig10}.
For the smaller values, $\epsilon=0.5$ and $\epsilon=1$, 
the overall behaviour is qualitatively similar
to that of Fig.~\ref{fig2}, showing 
a crossover from power law (with disorder dependent exponent) to logarithmic behavior, as
in Eq.~(\ref{num.7}). The data for the highest disorder value $\epsilon=2$, instead, display
a novel behaviour. The preasymptotic power law regime disappears and is replaced by a pronounced peak in the effective
exponent, which reveals strong pinning of the interfaces. Therefore, the nature of the crossover
is qualitatively different for {\it weak} and for {\it strong} disorder.
The detailed investigation of this novel feature is delayed to a future publication.

\begin{figure}
\centering
\rotatebox{0}{\resizebox{.85\textwidth}{!}{\includegraphics{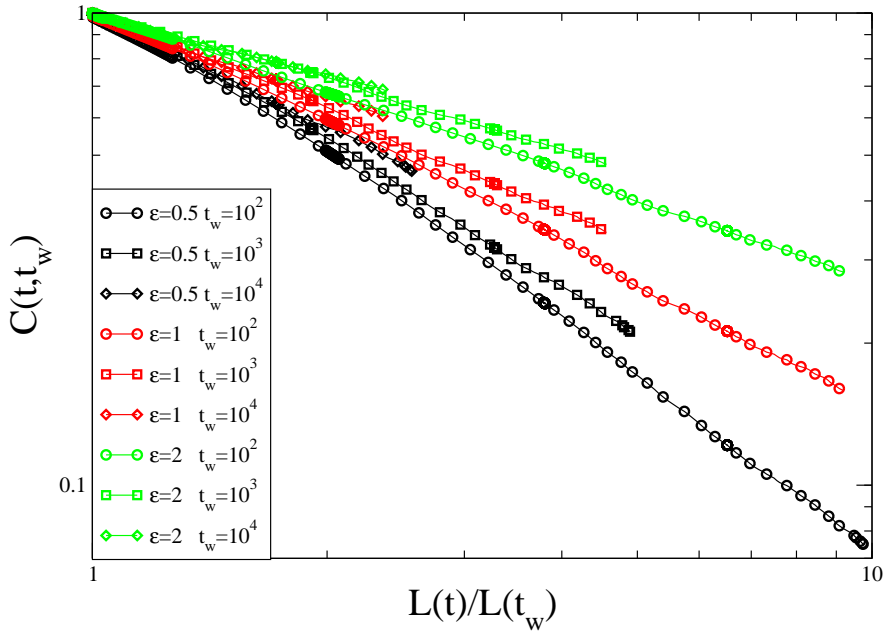}}}
\caption{(Color online) Autocorrelation function in $d=3$ for different 
$\epsilon$ and $t_w$ values.}
\vspace{1cm}
\label{fig11}
\end{figure}

The data for the autocorrelation function are displayed in Fig. \ref{fig11}. 
As stated above,
the quality of the data is not sufficient to carry out a precise scaling analysis as in Sec. \ref{s4}. 
In particular, it is not possible to extract the characteristic length $\lambda$ reliably
and to organise the plot with a choice of $t_w$ values aimed to keep constant the ratio
$L_w/\lambda$.
However, the evident $\epsilon$-dependence in Fig. \ref{fig11} is quite sufficient
to exclude SU validity. 

As a final remark, it should be noted that with the algorithm illustrated in Sec.~\ref{SD}
we consider a quench to $T=0$, while in Refs.~\cite{rao,aron} quenches to $T>0$ were considered.
Therefore, it remains an open question, to be investigated, whether the discrepancy
between ours and previous results may be related to the role of the final temperature.

\section{Summary and Discussion}
\label{s6}

Let us conclude this paper with a summary and discussion of our results. We have recently initiated a 
large-scale simulation study of nonconserved domain growth in disordered systems. Our study was 
motivated by some ambiguities existing in the available literature: \\
(a) The precise nature of the asymptotic growth law in the standard models (RBIM, RFIM, etc.) was unclear. \\
(b) Quantities like the equal-time correlation function (or its Fourier transform, the structure factor) 
showed SU, i.e., the scaling functions were independent of disorder. It was not clear to us why 
the crossover in the domain growth law was not accompanied by a corresponding crossover in the correlation function. \\
(c) There were very few studies of two-time quantities like the autocorrelation function and the response function.

With this background, we investigated two-time quantities in the $d=1,2$ RBIM \cite{EPL,CLMPZ} and, with the
present paper, we have extended the investigation to the $d=2,3$ RFIM with Glauber spin-flip kinetics. 
Our results can be summarized as follows: \\
(i) First, we have formulated a general scaling framework for the study of disordered domain growth. 
The framework is based on the RG concept of asymptotically relevant parameter, and proved very convenient for 
interpreting our earlier RBIM results \cite{CLMPZ}. In this paper, we have used this framework 
to successfully understand crossovers in the RFIM. \\
(ii) Second, we find that there is a crossover in the domain growth law from a preasymptotic regime showing 
power law growth with a disorder dependent exponent [$L(t) \sim t^{1/\overline{z}(\epsilon)}$] to 
an asymptotic regime with logarithmic growth [$L(t) \sim (\ln t)^{1/\varphi}$ with $\varphi \simeq 1.5$]. 
Following the analysis of Ref.~\cite{ppr04}, this can be related to an underlying crossover 
from logarithmic dependence
of the free energy barriers on the domain size to power law dependence. The mechanism producing
the power law dependence is particularly clear in $d=1$, where the interface motion
can be mapped into the random walk in a random potential of the Sinai type~\cite{decandia,monthus}.\\
(iii) Third, and perhaps most important, we find that the autocorrelation function does not obey SU. 
The scaling function shows a crossover corresponding to the crossover in the growth law.

In the light of the above results, what is the path ahead? We are currently investigating other 
disordered systems to confirm whether the domain growth scenario is consistent with the scaling 
picture developed here. It is also relevant to re-examine earlier results demonstrating SU for 
the equal-time correlation function. It is possible that the equal-time correlation function has a delayed crossover, 
and may violate SU in larger and longer simulations. Alternatively, it could be that the 
equal-time correlation function is a relatively crude and feature-less characteristic of the morphology. 
It may be worthwhile to study more sophisticated measures of the morphology \cite{bpc96}, 
which could show differences between pure and disordered phase ordering systems. 
Our studies demonstrate that there remain many unanswered questions in this area. 
We hope that our work will motivate fresh interest in these problems.

\end{document}